\documentclass{elsart}
\usepackage{epsf}
\newcommand{\ro}{\mbox{\boldmath$\rho$}}
\newcommand{\bpi}{\mbox{\boldmath$\pi$}}
\newcommand{\bn}{\mbox{\boldmath$\nabla$}}
\begin{document}
\runauthor{Dzyubenko}
\begin{frontmatter}
\title{{\small\ To be published in {\it Solid State Commun.} {\bf 113}, 
No.~12, p.~683 (2000).}\\[6pt]
Two-dimensional charged electron-hole complexes
          in magnetic fields: \\
  Keeping magnetic translations preserved}
\author[TP,GPI]{A. B. Dzyubenko}
\address[TP]{Institut f\"{u}r Theoretische Physik, J.W.
             Goethe-Universit\"{a}t, \\
               60054 Frankfurt,  Germany}
\address[GPI]{General Physics Institute, Russian Academy of Sciences, \\
               Moscow 117942, Russia}
\begin{abstract}
Eigenstates of two-dimensional charged electron-hole complexes
in magnetic fields are considered.
The operator formalism that allows one to partially separate
the center-of-mass motion from internal degrees of freedom is presented.
The scheme using magnetic translations is developed for calculating in
strong magnetic fields the eigenspectra of negatively charged excitons
$X^-$, a bound state of two electrons and one hole.
\end{abstract}
\begin{keyword}A. Semiconductors; A. Quantum wells; 
D. Electron-electron interactions
\end{keyword}
\end{frontmatter}

\section{Introduction}
\typeout{SET RUN AUTHOR to \@runauthor}

Identification \cite{Exp} of charged excitons in magneto-optical
spectra of quasi-two-dimensional (quasi-2D) systems has induced much
interest in the behavior of these three-particle electron-hole
($e$--$h$) complexes.
The negatively, $X^-$, and positively, $X^+$, charged excitons are
the bound states of two electrons and one hole ($2e$--$h$)
and two holes and one electron ($2h$--$e$), respectively.
In magnetic fields $B$, in addition to the spin-singlet, higher-lying
triplet states of $X^-$ and $X^+$ have been observed \cite{Exp}.
Theoretically, free charged excitons have been studied
in strictly 2D systems in the limit of high \cite{AHM}
and low \cite{Stebe} magnetic fields
and in quasi-2D systems  at high magnetic fields \cite{Chap,Whit97}.
For one-component electron systems in magnetic fields,
the center-of-mass motion separates from internal degrees
of freedom. The well-known Kohn theorem \cite{Kohn}, which states
that the electron cyclotron resonance is not shifted
or broadened by electron-electron interactions,
is based on this fact. For $e$--$h$ systems such a complete separation
is not possible in magnetic fields. Nonetheless,
any charged interacting system in a uniform $B$ possesses an
exact dynamical symmetry --- magnetic translations
(\cite{Simon,Hirsch} and references therein).
It has been shown recently \cite{Dz99} that due to this symmetry,
magneto-optical transitions of charged $e$--$h$ complexes are
governed by an exact selection rule, which leads to some rather
unexpected spectroscopic consequences for charged excitons in $B$.
In this work, using an operator formalism,
we construct a basis compatible with the exact dynamical symmetries
--- rotations about the $B$-axis and magnetic translations.
Physically, this is equivalent to a partial separation of the
center-of-mass motion from internal degrees of freedom in $B$
\cite{Simon,Hirsch}.
We demonstrate that this basis can be used for high-accuracy
and rapidly convergent calculations
of bound $X^-$ states in strong magnetic fields.
Our results can also be relevant for atomic ions with not too
large mass ratios in ultrastrong magnetic fields \cite{Hirsch}.

\section{Basis compatible with magnetic translations}

   We consider a strictly 2D system containing two electrons and one
hole in a perpendicular magnetic field ${\bf B}=(0,0,B)$
described by the Hamiltonian
\begin{eqnarray}
                \label{H} 
  H &=& H_0 + H_{\rm ee} + H_{\rm eh} \, , \\
                \label{H_0} 
  H_0 &=& \sum_{i=1,2} \frac{\hat{\bpi}_{ei}^2}{2m_e} +
                      \frac{\hat{\bpi}_{h}^2 }{2m_h} \, , \\
            \label{H_int} 
  H_{\rm ee} &=& \frac{e^2}{\epsilon |{\bf r}_1-{\bf r}_2|}
             \quad  ,  \quad
  H_{\rm eh} = -\sum_{i=1,2} \frac{e^2}
                                  {\epsilon |{\bf r}_i-{\bf r}_h|} \, ,
\end{eqnarray}
where $\hat{\bpi}_j = -i\hbar \bn_j -
 \frac{e_j}{c} {\bf A}({\bf r}_j)$
are kinematic momentum operators.
We will use the symmetric gauge
${\bf A} = \frac12 {\bf B} \times {\bf r}$.
The exact eigenstates can be characterized by the total angular momentum
projection $M_z$, an eigenvalue of
$\hat{L}_z=\sum_j ({\bf r}_j \times -i\hbar\bn_j)_z$,
by the total spin of two electrons
$S_e=0$ (singlet states) or $S_e=1$ (triplet states),
and the spin state of the hole $S_h$. The latter simply factors out
and will be disregarded.
Performing an orthogonal transformation of the coordinates
$\{ {\bf r}_1,{\bf r}_2,{\bf r}_{h}\} \rightarrow
 \{ {\bf r},{\bf R}, {\bf r}_h \}$,
where ${\bf r} = ({\bf r}_1 - {\bf r}_2)/\sqrt{2}$ is the electron
relative and ${\bf R} = ({\bf r}_1 + {\bf r}_2)/\sqrt{2}$
center-of-mass coordinates,
the complete orthonormal basis with a fixed value of $M_z$
can be constructed \cite{Dz_PLA} (see also \cite{Dz_PLA92})
as an expansion in Landau levels (LL's)
\begin{equation}
        \label{basis_M}
  \phi^{(e)}_{n_1 m_1}({\bf r}) \,
  \phi^{(e)}_{n_2 m_2}({\bf R}) \,
  \phi^{(h)}_{n_{h}m_{h}}({\bf r}_{h}) \, .
\end{equation}
Here $\phi^{(e)}_{n m}({\bf r})=\phi^{(h)*}_{n m}({\bf r})$
are the $e$- and $h$- single-particle factored wave functions in $B$;
$n$ is the LL quantum number and $m$ is the oscillator quantum number
(see, e.g., \cite{Simon,Hirsch}). For, e.g., zero LL's
\begin{equation}
              \label{zero_LL}
   \phi^{(e)*}_{0m}({\bf r}) = \phi^{(h)}_{0m}({\bf r}) =
   \frac{1}{(2\pi m!\ell_B^2)^{1/2}}
   \left( \frac{z}{\sqrt{2}\ell_B} \right)^m
   \exp\left(-\frac{r^2}{4\ell_B^2} \right)  \, ,
\end{equation}
where $ z = x + iy$ is the 2D complex coordinate and
$\ell_B=(\hbar c/eB)^{1/2}$.
The factored wave functions are constructed with the help of the
oscillator Bose ladder operators: For electrons (the charge $-e<0$)
\begin{equation}
              \label{e_LL}
\phi^{(e)}_{n m}({\bf r})=\frac{1}{\sqrt{n!m!}}
\langle {\bf r} |(A^{\dag}_{e})^n (B^{\dag}_{e})^m |0\rangle \, ,
\end{equation}
here the intra-LL operators
$B^{\dag}_{e}({\bf r}_j) = -i \sqrt{c/2 \hbar B e} \, \hat{ K }_{j-}$,
where $\hat{ K }_{j\pm}= \hat{ K }_{jx} \pm i \hat{ K }_{jy}$
and
$\hat{\bf K}_j =
\hat{\bpi}_j - \frac{e_j}{c} {\bf r}_j \times {\bf B}$ (see, e.g.,
\cite{Simon,Hirsch}).
The electron inter-LL operators are
$A^{\dag}_{e}({\bf r}_j) = -i \sqrt{c/2 \hbar B e} \, \hat{ \pi }_{j+}$,
where $\hat{ \pi }_{j\pm}= \hat{ \pi }_{jx} \pm i \hat{ \pi }_{jy}$.
The operators commute as $[A_{e},A^{\dag}_{e}]=1$,
$[B_{e},B^{\dag}_{e}]=1$,
and $[A_{e},B^{\dag}_{e}]=[A_{e},B_{e}]=0$.
The analogous intra-LL and inter-LL operators for the hole
(the charge $e>0$) are, respectively,
$B^{\dag}_{h}({\bf r}_h) = -i \sqrt{c/2 \hbar B e} \, \hat{ K }_{h+}$
and
$A^{\dag}_{h}({\bf r}_h) = -i \sqrt{c/2 \hbar B e} \, \hat{ \pi }_{h-}$.
These can be considered as linear functions of spatial coordinates and
derivatives and have the form
\begin{eqnarray}
        \label{lad_en}
    A^{\dag}_{e}({\bf r}) = B^{\dag}_{h}({\bf r}) &=& \frac{1}{\sqrt{2}}
  \left( \frac{z}{2\ell_B} - 2\ell_B \frac{\partial}{\partial z^{\ast}}
  \right) \, , \\
      \label{lad_em}
    B^{\dag}_{e}({\bf r}) = A^{\dag}_{h}({\bf r}) &=& \frac{1}{\sqrt{2}}
  \left(\frac{z^{\ast} }{2\ell_B}-2\ell_B\frac{\partial}{\partial z }
  \right) \, .
\end{eqnarray}
Single-particle angular momentum projection operators
$\hat{L}_{ze}=A^{\dag}_{e}A_{e} -  B^{\dag}_{e}B_{e}$
and $\hat{L}_{zh}=B^{\dag}_{h}B_{h} - A^{\dag}_{h}A_{h}$, so that
$m_{ze}=-m_{zh}=n-m$.
The basis (\ref{basis_M}) includes therefore different three-particle
$2e$--$h$ states such that $M_z= n_1+n_2-m_1-m_2-n_h+m_h$ is fixed.
Permutational symmetry of identical particles requires
that for electrons in the spin-singlet $S_e=0$ (triplet $S_e=1$) state
the relative motion angular momentum $n_1-m_1$ should be even (odd).
The basis (\ref{basis_M}) proved to be effective in strong $B$ for
studying impurity-bound states of $e$--$h$ complexes \cite{Dz_PLA},
collective excitations --- magnetoplasmons and spin-waves \cite{Dz&L93},
and effects of lateral confinement in quantum dots in $B$
\cite{Xdot,Hawrylak}. The equivalent LL expansion (using the coordinates
$\{ {\bf r}_1,{\bf r}_2,{\bf r}_{h}\}$) has been exploited
\cite{Chap,Whit97} for studying {\em free\/} charged excitons in $B$.
However, for translationally invariant systems the basis (\ref{basis_M})
is {\em not compatible\/} with the magnetic translations.

  Indeed, the Hamiltonian (\ref{H}) commutes with the operator of
the magnetic translations $\hat{\bf K} = \sum_{j} \hat{\bf K}_j$
\cite{Simon,Hirsch}.
Noting that $[\hat{K}_x, \hat{K}_y] = - i \frac{\hbar B}{c} Q$,
where the total charge $Q \equiv \sum_j e_j=-e$ for the $X^-$,
one obtains the lowering and raising Bose
ladder operators for the {\em whole system\/} \cite{Simon,Hirsch,Dz99}
\begin{equation}
  \label{kmin}
  \hat{k}_{\pm} = \pm \frac{i}{\sqrt{2}} (\hat{k}_x  \pm i \hat{k}_y)
            \quad , \quad
    [\hat{k}_{+}, \hat{k}_{-}]=-\frac{Q}{|Q|}=1 \, ,
\end{equation}
here $\hat{{\bf k}} = \sqrt{c/\hbar B |Q|} \, \hat{\bf K}$.
Therefore,
$\hat{{\bf k}}^2 = \hat{k}_{+} \hat{k}_{-} + \hat{k}_{-} \hat{k}_{+}$
has the discrete oscillator eigenvalues $2k+1$, $k=0, 1, \ldots$.
These can be used, together with $M_z$,
for labelling of exact charged eigenstates of (\ref{H}).
Due to the non-commutativity of
$\hat{K}_x$ and  $\hat{K}_y$, there is the macroscopic Landau
degeneracy in $k$. Note now that
$\hat{{\bf k}}^2 = \left( \sum_j \hat{{\bf k}}_j \right)^2=
\sum_j \hat{{\bf k}}^2_j +
\sum_{i \neq j} \hat{{\bf k}}_i \cdot \hat{{\bf k}}_j$
is {\em not diagonal\/} in the basis (\ref{basis_M}) due to the cross
terms $\sum_{i \neq j} \hat{{\bf k}}_i \cdot \hat{{\bf k}}_j$.

In order to make the basis (\ref{basis_M}) compatible with the magnetic
translations, a canonical transformation diagonalizing $\hat{{\bf k}}^2$
should be performed. We deal formally with a set of coupled harmonic
oscillators. Note then that
\begin{equation}
        \label{k-}
     \hat{k}_{-} =
      B_e^{\dag}({\bf r}_1) + B_e^{\dag}({\bf r}_2) - B_h({\bf r}_h)
                 = \sqrt{2}\,B_e^{\dag}({\bf R}) - B_h({\bf r}_h)
\end{equation}
and define
\begin{equation}
        \label{Bog1}
 \tilde{B}_e^{\dag}({\bf R}) =
                 u B^{\dag}_{e}({\bf R}) - v B_{h}({\bf r}_h)
                     \quad  ,  \quad
  \tilde{B}_e({\bf R})  =
                 u B_{e}({\bf R}) - v B^{\dag}_{h}({\bf r}_h)  \, ,
\end{equation}
where $u= \sqrt{2}$, $v=1$.
It is this pair of Bose ladder operators in which $\hat{{\bf k}}^2$ is
diagonal:
$\hat{{\bf k}}^2 = 2\tilde{B}_e^{\dag} \tilde{B}_e  + 1$.
Equation (\ref{Bog1}) is in fact a Bogoliubov canonical transformation
$\tilde{B}_e^{\dag} = S B_e^{\dag} S^{\dag}$
generated by the unitary operator
(see, e.g., \cite{Kirzhnits,Wegner})
\begin{equation}
        \label{BogS}
  S= \exp \{ \Theta [B_e({\bf R})B_h({\bf r}_h) -
                     B^{\dag}_{h}({\bf r}_h) B^{\dag}_{e}({\bf R}) ] \}
\end{equation}
with $u= {\rm ch} \Theta=\sqrt{2}$, $v= {\rm sh} \Theta =1$.
The second pair of linearly independent transformed operators
$\tilde{B}_h^{\dag}({\bf r}_h) = S B^{\dag}_{h}({\bf r}_h) S^{\dag}$
and
$\tilde{B}_h({\bf r}_h) =  S B_{h}({\bf r}_h) S^{\dag}$ are
\begin{equation}
        \label{BogS_2}
 \tilde{B}_h^{\dag}({\bf r}_h) =
                      u B^{\dag}_{h}({\bf r}_h) - v B_{e}({\bf R})
                     \quad  ,  \quad
 \tilde{B}_h({\bf r}_h) =
                      u B_{h}({\bf r}_h) - v B^{\dag}_{e}({\bf R}) \, .
\end{equation}
The complete orthogonal basis compatible with both axial and
translational symmetries therefore is
\begin{equation}
        \label{basis}
       A_e^{\dag}({\bf r})^{n_1}
       A_e^{\dag}({\bf R})^{n_2}
       A_h^{\dag}({\bf r}_h)^{n_h}
   \tilde{B}_e^{\dag}({\bf R})^k
           B_e^{\dag}({\bf r})^m
   \tilde{B}_h^{\dag}({\bf r}_h)^l |\tilde{0} \rangle \, .
\end{equation}
In (\ref{basis}) the oscillator quantum number is fixed and equals $k$
while $M_z= -k -m + l + n_1 + n_2- n_h$ and
$n_1-m$ is even (odd) for  $S_e=0$ ($S_e=1$).
The Hamiltonian (\ref{H}) is block-diagonal in the quantum numbers
$k,M_z,S_e$. Moreover, due to the Landau degeneracy in $k$,
it is sufficient to consider the $k=0$ states only.
This effectively removes one degree of freedom and corresponds
to a partial separation of the center-of-mass motion from internal
degrees of freedom for a charged $e$--$h$ system in a magnetic field
(cf.\ \cite{Simon,Hirsch}).

In (\ref{basis}) the new vacuum $|\tilde{0} \rangle = S |0\rangle$
has been introduced. Disentangling the operators in the exponent
of $S$ (see, e.g., \cite{Kirzhnits,Wegner}), one obtains
\begin{eqnarray}
        \label{Sexp}
  S & = & \exp \left( - {\rm th} \Theta \, B^{\dag}_{h} B^{\dag}_{e}
               \right) \\
    \nonumber
   && \times \exp \left( -\ln({\rm ch} \Theta )
          [B^{\dag}_{e} B_{e} + B^{\dag}_{h} B_{h} +1 ] \right)
     \exp \left( {\rm th} \Theta \, B_eB_h \right) \, ,
\end{eqnarray}
so that
\begin{equation}
 |\tilde{0} \rangle = S |0\rangle = \frac{1}{{\rm ch} \Theta }
  \exp\left( - {\rm th}\Theta \,
  B^{\dag}_{h}({\bf r}_h) B^{\dag}_{e}({\bf R}) \right) |0\rangle \, .
\end{equation}

For a charged system of $N_e$ electrons and $N_h$ holes
(with, e.g., $N_e>N_h$),
a transformation analogous to (\ref{Bog1})--(\ref{BogS_2}) can also be
performed. It should involve the intra-LL $e$- and $h$- center-of-mass
operators $B^{\dag}_{e}({\bf R}_e)$ and $B_{h}({\bf R}_h)$
with ${\rm th} \Theta = \sqrt{N_h/N_e}$;
here ${\bf R}_e = \sum_{i=1}^{N_e} {\bf r}_{ei}/\sqrt{N_e}$
and ${\bf R}_h = \sum_{j=1}^{N_h} {\bf r}_{hj}/\sqrt{N_h}$.

\section{$X^-$ states in lowest Landau levels}

  We now demonstrate how the developed formalism works. We will
consider the limit of high magnetic fields \cite{AHM,Whit97,Dz99}
\begin{equation}
      \label{highB}
 \hbar\omega_{\rm ce} \, , \,  \hbar\omega_{\rm ch} \, , \,
|\hbar\omega_{\rm ce} - \hbar\omega_{\rm ch}| \gg
E_0 = \sqrt{\frac{\pi}{2}} \, \frac{e^2}{\epsilon l_B} \, ,
\end{equation}
when mixing between different LL's can be neglected.
$E_0$ is the characteristic energy of the Coulomb interactions
in strong $B$, $\hbar\omega_{\rm ce(h)}=\hbar eB/m_{\rm e(h)}c$.
Charged magnetoexcitons can then be labeled by the total electron
LL number $n_e=n_1+n_2$ and by the hole LL number $n_h$.
Indeed, when (\ref{highB}) is fulfilled, the states having different
quantum numbers $n_en_h$  and $n'_en'_h$ are only weakly
$\sim E_0/|(n'_e-n_e)\hbar\omega_{\rm ce} +
           (n'_h-n_h)\hbar\omega_{\rm ch}|$
mixed by the Coulomb interactions \cite{Dz_PLA92}.

We focus on the states in zero LL's [$n_1=n_2=n_h=0$ in (\ref{basis})].
The operators  (\ref{Bog1}), (\ref{BogS_2})
have a simple representation in the new coordinates
$\ro_1 = \sqrt{2} \, {\bf R} - {\bf r}_h$
and
$\ro_2 = \sqrt{2} \, {\bf r}_h - {\bf R}$:
$\tilde{B}_e^{\dag}({\bf R}) =   B_e^{\dag}(\ro_1)$
and
$\tilde{B}_h^{\dag}({\bf r}_h) =  B_h^{\dag}(\ro_2)$.
The complete infinite orthonormal basis in zero LL's with fixed $k=0$
and arbitrary $M_z=l-m$ takes the form
\begin{equation}
        \label{bas_k}
       \frac{1}{(m!l!)^{1/2}}
        B_e^{\dag}({\bf r })^m B_h^{\dag}(\ro_2)^l
            |\tilde{0} \rangle \equiv | m \, l \rangle \,
\end{equation}
with odd $m=2p+1$ (even $m=2p$), $p=0,1, \ldots$
in the electron triplet $S_e=1$ (singlet $S_e=0$) states.
The Coulomb interactions in the new variables are
\begin{equation}
        \label{Ham_rho}
H_{\rm ee}= \frac{e^2}{\sqrt{2} \epsilon r} \quad , \quad
H_{\rm eh}= - \frac{\sqrt{2}e^2}{\epsilon|\ro_2 - {\bf r}|}
            - \frac{\sqrt{2}e^2}{\epsilon|\ro_2 + {\bf r}|} \, .
\end{equation}
The matrix elements of the $e$--$e$ interaction
are diagonal in the basis (\ref{bas_k}):
\begin{equation}
        \label{mat_ee}
 \langle m_2 l_2 | H_{\rm ee} | m_1 l_1 \rangle =
 \delta_{m_1,m_2} \delta_{l_1,l_2} \frac{V_{0,m_1}}{\sqrt{2}}
         \quad , \quad
   V_{0,m}= \frac{(2m-1)!!}{2^mm!} \, E_0 \, ,
\end{equation}
where $V_{0,m}$ is the interaction of the electron with a fixed
negative charge $-e$ in zero LL (e.g., \cite{Dz_PLA}).
Due to the permutational symmetry, the two terms in $H_{\rm eh}$ give
the same contributions; calculations, however, are not so
straightforward as (\ref{mat_ee}).
This is connected with the fact that the coordinate transformation
$\{ {\bf r},{\bf R},{\bf r}_h \} \rightarrow
 \{ {\bf r},\ro_1,\ro_2 \}$ is {\em not orthogonal\/}.
As a result, the coordinate representation of the new vacuum
is not factored in $\ro_1$ and $\ro_2$:
\begin{equation}
   \langle {\bf r}\ro_1\ro_2| \tilde{0} \rangle =
    \frac{1}{\sqrt{2}\,(2\pi \ell_B^2)^{3/2}}
     \exp \left( -\frac{r^2+\rho_1^2+\rho_2^2+
      \sqrt{2} Z_1 Z_2^{\ast} }{4 \ell_B^2} \right) \, ,
\end{equation}
here $Z_j= \rho_{jx} +i\rho_{jy}$, $j=1,2$.
Therefore, when acting on the vacuum $|\tilde{0} \rangle$,
the operator
$B_h^{\dag}(\ro_2)=\tilde{B}_h^{\dag}({\bf r}_h)$
gives a combination
$(Z_2 + \frac{1}{\sqrt{2}}  Z_1)/\sqrt{2}\ell_B= z_h/2\ell_B$.
To eliminate the coordinate $\ro_1$, we perform the shift
$\ro_1 \rightarrow \tilde{\ro} =
\ro_1 + \frac{1}{\sqrt{2}} \ro_2=\frac{1}{\sqrt{2}}{\bf R}$
and obtain
\begin{eqnarray}
        \label{mat_eh}
 \langle m_2 l_2 | H_{\rm eh} | m_1 l_1 \rangle   &=&
 \int\! \frac{d^2 {\rho}_2}
             {2\cdot2\pi\ell_B^2\sqrt{2^{l_1+l_2}l_1!l_2!}}
  \exp\left( -\frac{{\rho}_2^2}{4\ell_B^2} \right)  \\
   \nonumber
 &&\times  \int\! d^2 r \,  \phi^{(e)*}_{0m_2}({\bf r})
         \frac{-2\sqrt{2}e^2}{ \epsilon | \ro_2 - {\bf r}|}
                  \phi^{(e)}_{0m_1}({\bf r})
\int\! \frac{d^2 \tilde{\rho}}{2\pi\ell_B^2}
  \exp\left(
    -\frac{\tilde{\rho}^2}{2\ell_B^2}\right)   \\
   \nonumber
  & & \times \left(\frac{\tilde{Z}^{\ast}}{\sqrt{2}\ell_B}  +
         \frac{Z_2^{\ast}}{2\ell_B} \right)^{l_2}
  \left(\frac{\tilde{Z}}{\sqrt{2}\ell_B} +
         \frac{Z_2}{2\ell_B}\right)^{l_1}
           \sim \delta_{l_1-m_1,l_2-m_2} \, .
\end{eqnarray}
Integrating out the variable $\tilde{\ro}$, we reduce
the problem to an effective {\em two-particle\/} $e$--$h$
problem in zero LL's (cf.\ \cite{Dz_PLA}).
The peculiarity of the situation is that the effective particles
are characterized by {\em different\/} magnetic lengths.
The matrix elements (\ref{mat_eh})
can be presented in the form ($m_1=m$, $m_2=m+s$, $l_1=l$, $l_2=l+s$)
\begin{equation}
         \label{mat_eh2}
 \langle m\!+\!s \, l\!+\!s | H_{\rm eh} | m \,l \rangle =
     (-2\sqrt{2}) 2^{-l-\frac{s}{2}} \sum_{k=0}^{l}
     \left( C_{l}^{k} C_{l+s}^{k+s} \right)^{\frac{1}{2}}
            U^{(\alpha=2)}_{km}(s) \, ,
\end{equation}
where $C_n^m$ are binomial coefficients
and the matrix elements of the Coulomb interparticle interactions in
zero LL's have been introduced:
\begin{eqnarray}
        \label{U_a}
   \nonumber
   \int\! d^2 r_1 &&\int\! d^2 r_2 \,
    \phi_{0m_2}^{(e)*}\left( {\bf r}_1 \right)
     \phi_{0k_2}^{(h)*}\left({\bf r}_2/\sqrt{\alpha} \right)
         \frac{e^2}{|{\bf r}_1 - {\bf r}_2|}
   \phi_{0k_1}^{(h)}\left( {\bf r}_2/\sqrt{\alpha} \right)
     \phi_{0m_1}^{(e)}\left( {\bf r}_1 \right)  = \\
 =&& \delta_{k_1-m_1,k_2-m_2}
  U^{(\alpha)}_{{\rm min}(k_1,k_2),{\rm min}(m_1,m_2)}(|m_1 - m_2|)
\end{eqnarray}
The matrix elements (\ref{U_a}) can be found analytically for arbitrary
$\alpha$:
\begin{eqnarray}
        \label{U_a_sum}
   U^{(\alpha)}_{mn}(s) & = & E_0
    \frac{\alpha^{\frac{s}{2}}
      \left[ m!(m+s)!n!(n+s)! \right]^{-\frac{1}{2}}}
       {(1+\alpha)^{s+\frac{1}{2}} \, 2^{m+n+s}}
   \sum_{k=0}^{m} \, \sum_{l=0}^{n} \,  C_m^k \, C_n^l  \\
     \nonumber
    &&   \times \frac{ \alpha^l }{ (1+\alpha)^{k+l} } \,
    [2(k+l+s)-1]!! \, [2(m-k) -1]!! \, [2(n-l)-1]!! \, .
\end{eqnarray}

Equations (\ref{mat_ee}), (\ref{mat_eh2}), and (\ref{U_a_sum})
determine the secular equation of the {\em infinite order\/}
that should be solved to obtain the three-particle $2e$--$h$
states in zero LL's. A truncation of the basis should naturally
be performed. An important property of the developed
basis (\ref{basis}) [and (\ref{bas_k})] is that such a truncation
{\em does not break\/} the translational invariance.
On the contrary, a truncation of the basis (\ref{basis_M}),
as performed in \cite{Chap,Whit97} (see also \cite{AHM}),
leads to spurious mixing of different $k$-states and violates the exact
magneto-optical selection rule \cite{Dz99} --- the conservation of the
oscillator quantum number $k$.

\begin{table*}
\caption{Singlet $X^-_{sn_en_h}$ and triplet $X^-_{tn_en_h}$
charged magnetoexcitons in Landau levels $n_en_h$}
\vspace*{.6cm}
\label{table}
\begin{center}
\begin{tabular}{c r c c }
\hline
             &{$M_z$}     & Interaction energy ($E_0$) &
                                        Binding energy ($E_0$) \\
\hline
$X^-_{t00}$  & -1$^{\dag}$  & -1.04345 & 0.04345              \\
$X^-_{s01}$  & -3$^\ddag$   & -0.78056 & 0.20690               \\
$X^-_{t01}$  & -4$^\ddag$   & -0.75776 & 0.18410               \\
$X^-_{t10}$  & 1$^{\dag}$   & -1.08596 & 0.08596               \\
\hline
\end{tabular}
\end{center}
\vspace*{0.6cm}
\noindent
$^{\dag}$ the only bound states in the given LL's $n_en_h$. \\
 $^\ddag$ the ground states among many other bound states with the same
 $n_en_h$.
\end{table*}

The developed approach also provides an effective computational tool:
First, we have been able to remove one degree of freedom in the
three-particle problem, so that configurational space is substantially
reduced (cf.\ \cite{Whit97}). As a result, with finite-size calculations
it is even possible to reproduce
with a reasonable accuracy the three-particle continuum --- a neutral
magnetoexciton plus a scattered electron \cite{Dz99}.
Second, for {\em bound\/} $X^-$ states lying outside the continua
we have extremely rapid convergence within each LL\@. This is associated
with the {\em exponential\/} decay of the off-diagonal matrix elements
(\ref{mat_eh2}). Consider, e.g., the $k=0$ triplet $X^-_{tn_e=0n_h=0}$
state in zero LL's with $M_z=-1$.
The asymptotic behavior of the relevant off-diagonal Coulomb $e$--$h$
matrix elements is
\begin{equation}
        \label{mat01}
  \langle 2s\!+\!1 \, 2s| H_{\rm eh} | 1 \, 0 \rangle =
  -\frac{2\sqrt{2}}{2^s} U^{(\alpha=2)}_{01}(2s)
  \simeq - \sqrt{\frac{32}{27\pi}} \left(\frac{1}{9}\right)^{s}E_0
 \: , \: s \gg 1 \, .
\end{equation}
Also, even the 1$\times$1 matrix Hamiltonian in the basis (\ref{bas_k})
$\langle 1\, 0 | H_{\rm ee}+ H_{\rm eh} | 1\, 0 \rangle = -1.0073E_0$
ensures the $X^-$ binding: it gives a positive binding energy
$0.0073E_0$; this is relative to the ground state energy $-E_0$ of
the neutral $X_{n_e=0n_h=0}$ magnetoexciton in zero LL's.
As a result, the $X^-_{t00}$ binding energy can be calculated with
virtually unlimited accuracy and equals $0.043452E_0$; this value is
compatible with \cite{AHM,Whit97}. Not accounting for the Landau
degeneracy in $k$, the $X^-_{t00}$ state with $M_z=-1$ is the only
low-lying bound $X^-$ state in zero LL's: there are no other bound
triplet or singlet states \cite{AHM,Whit97,Dz99}.

\begin{figure*}[t]
\epsfxsize=5.1in
\vspace*{0.6cm}
\epsffile{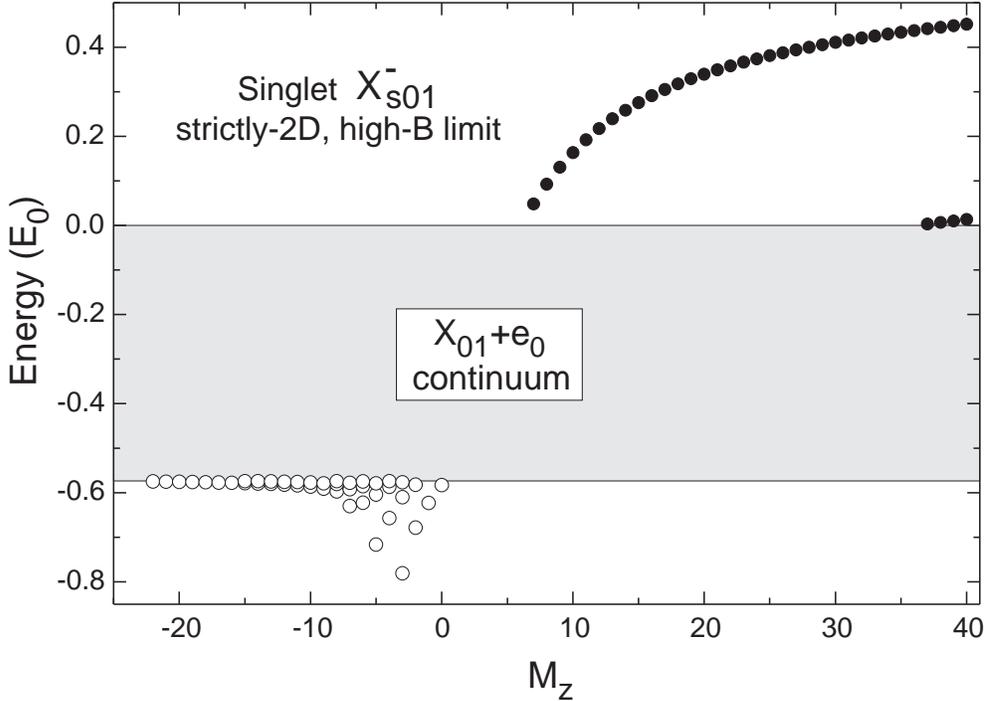}
\caption{Bound and scattering three-particle $2e$--$h$
singlet $S_e=0$ states in the ($n_en_h$)=(01) LL's.
Note that only the $k=0$ states are shown.
The energy is given relative to the energy of free LL's
$\frac{1}{2}\hbar\omega_{\rm ce} + \frac{3}{2}\hbar\omega_{\rm ch}$,
in units of $E_0= \protect\sqrt{\pi/2} \, e^2/\epsilon l_B$.
Open circles correspond to low-lying $X^-_{s01}$ states.
Filled circles correspond to excited states originating from bound
internal motion of two electrons in 2D in high magnetic fields.
The shaded area shows the continuum corresponding to the
neutral magnetoexciton $X_{01}$ plus a scattered electron in the
$n_e=0$ LL; the lower continuum edge lies at an energy $-0.5737E_0$.
}
\label{fig1}
\end{figure*}

Similar considerations apply to the $X^-$ states in higher LL's.
Some of the results for the $X^-$ ground states are presented in
Table~1. There is only one bound $X^-$ state
in the first electron LL (the basis (\ref{basis}) includes the states
with $n_1=1$, $n_2=0$, $n_h=0$ and $n_1=0$, $n_2=1$, $n_h=0$).
This state is the triplet $X^-_{t10}$ with $M_z=1$, whose binding energy
is almost twice that of the $X^-_{t00}$ state in zero LL's \cite{Dz99}.
This resembles a stronger binding of the triplet $D^-$
state (two electrons bound by a donor ion)
in the first electron LL \cite{Dz_PLA92} and has the same
physical origin.
The $X^-_{t10}$ binding energy is counted from the lowest possible
unbound state in the same LL's, which is the neutral magnetoexciton
$X_{n_e=0n_h=0}$
with the second electron in the scattering state in the $n_e=1$ LL.
As calculations show, there are {\em many\/} bound $X^-$
states in the next hole LL [$n_1=n_2=0$, $n_h=1$ in (\ref{basis})] ---
both triplets $X^-_{t01}$ and singlets $X^-_{s01}$ (see Fig.~1).
These are lying below the ground state of the
neutral magnetoexciton $X_{n_e=0n_h=1}$; the latter has the energy
$-0.57366E_0$.
Due to this small
binding energy of the neutral $X_{n_e=0n_h=1}$ magnetoexciton
(comparatively to the $X_{n_e=0n_h=0}$ magnetoexciton),
the triplet $X^-_{t01}$ and singlet $X^-_{s01}$ ground states
have rather large binding energies (Table~1).
In all LL's, there are also higher-lying bound three-particle $2e$--$h$
states \cite{Dz99} originating from the internal bound motion of 2D
electrons in strong magnetic fields.
These states appear in the spectrum at relatively large positive values
of the total $M_z$, that correspond to the hole being
at large distances from the electrons
(cf.\ with the similar states in the $D^-$ problem  \cite{Dz_PLA92}).

\section{Summary}

In conclusion, we have developed a formalism that allows
one to preserve the exact symmetry --- magnetic translations ---
when performing the Landau level expansion for charged electron-hole
complexes in magnetic fields. This is achieved by using
the Bogoliubov canonical transformation mixing the center-of-mass
motions of the electron and hole subsystems.
The effectiveness of the scheme has been demonstrated for
high-accuracy and rapidly convergent calculations
of two-dimensional charged excitons $X^-$ in magnetic fields.
This can be useful for studying the eigenspectra of
charged excitons in quasi-two-dimensional quantum wells at strong
and intermediate magnetic fields.

\section*{Acknowledgments}
The author is grateful to H.\ Haug and A.Yu.\ Sivachenko
for useful discussions. This work was supported
by the Humboldt Foundation and the grants RBRF~97-2-17600 and
``Nanostructures''~97-1072.

\end{document}